\documentclass[a4paper,11pt]{article}
\usepackage[english,spanish]{babel}
\usepackage{amsfonts} 
\usepackage{amssymb}  
\usepackage{graphicx} 
\usepackage{amsmath,amsthm}  
\usepackage[utf8]{inputenc}
\usepackage{latexsym}
\usepackage{geometry}
\usepackage{color}
\usepackage{tikz}
\def\email#1{\it #1}

\begin{document}

\title{Condensation of neutral vector bosons with magnetic moment}%
\author{Gretel Quintero Angulo \\ 
Facultad de F{\'i}sica, Universidad de la Habana,\\ San L{\'a}zaro y L, Vedado, La Habana 10400, Cuba\\
\email{gquintero@fisica.uh.cu} \\[5pt]
Aurora Pérez Mart\'{\i}nez \\
Departamento de Física Teórica, Instituto de Cibern\'{e}tica Matem\'{a}tica y F\'{\i}sica,\\	Calle E esq 15 No. 309, Vedado, La Habana 10400, Cuba\\
Instituto de Ciencias Nucleares, Universidad Nacional Autónoma de México,\\	A. P.70-543, 04510 C. México, México \\
\email{aurora@icimaf.cu}\\[5pt]
Hugo Pérez Rojas \\
Departamento de Física Teórica, Instituto de Cibern\'{e}tica Matem\'{a}tica y F\'{\i}sica,\\ Calle E esq 15 No. 309, Vedado, La Habana 10400, Cuba \\
\email{hugo@icimaf.cu}
}%
\maketitle
\thanks{}%
\date{}%

\begin{abstract}
We study the equation of motion of neutral vector bosons bearing a magnetic moment (MM). The effective rest mass of vector bosons is a decreasing function of the magnetic field intensity. Consequently a  diffuse condensation of the  bosons appears below a critical value of the field. For typical values of densities and magnetic fields the magnetization is positive  and the neutral boson system  can  maintain a magnetic field self-consistently.   A discussion of the relevance in astrophysics is presented.
\end{abstract}


\section{Introduction}	

There is a diversity of structures associated with a wide range of magnetic fields ($10^{-9}-10^{15}~G$) cohabiting in our Universe. Salient examples of magnetized structures are galaxies (radius $1.5 \times 10^{18}~km$) and compact objects (radius $20~km$). Although some theories have been proposed to explain the origin of such magnetic fields, this issue is far from being exhausted and it is still under great debate.


As it is well known, the magnetic field modifies mainly the behavior of matter at the microscopic scale, but there are microscopic effects that lead to a sensitive variation in the macroscopic features.

The actual rate of expansion of our Universe and the light element abundances depend on the magnetic field and put limit to its values at the early universe\cite{Grasso199673}.  On the other hand the size and shape of the compact object depend on the magnetic field too\cite{0954-3899-36-7-075202}. There are also some phenomena at Astrophysical scale without explanation, as kicks and jets of pulsars, and magnetic fields could contribute to explain them\cite{MNL2:MNL2848,Charbonneau:2009ax}. Furthermore, a magnetic field breaks the rotational symmetry and it is reflected not only in the particle spectra but also in the energy momentum tensor of the system. The latter becomes anisotropic which imply the splitting of the parallel and perpendicular pressures along the magnetic field.



The aim of this work is to study the thermodynamical properties of a neutral spin one boson gas bearing a magnetic moment in presence of a constant magnetic field. Magnetization and Bose-Einstein condensation are discussed. Neutral vector bosons can be mesons and other  paired fermions leading to total integer spin.  These particles could be relevant components/participants of astrophysical objects/phenomena. In particular, we are pursuing some insights on the nature of jets -usually a stream of matter ejected from a given compact object along its axis of rotation.

We will  concern with the phenomenology of a neutral vector boson gas with spin one bearing a MM in a magnetized medium  disregarding the realistic conditions in which it may be realized. In particular we consider a positronium gas, characterized by a  mass approximately  $2m_e$, ($m_e$ is the electron mass) and a twice  the electron magnetic moment ($\kappa =2\mu_B$ being $\mu_B$  the Bohr magneton).

In Ref\cite{Elizabeth} has been discussed the properties of a  charged vector bosons gas with the aim to build a model of jets. As we will see below the ground state of neutral bosons  has a similar form that the charged one discussed in\cite{Elizabeth}. However the thermodynamical quantities for ground state difieres in the neutral case from the charged one. In the latter case, the momentum component perpendicular to the magnetic field -$p_{\perp}$- is quantized and the density of states becomes proportional to $eB$ ($2\int d^3p \rightarrow \sum_n (2-\delta_{n0}) \frac{eB}{(2\pi)^2}\int dp_3$). In the neutral case there is not quantization in any component of the momentum and the density of states is not proportional to the magnetic field.

Our paper is organized as follows. In Section~\ref{eom}, we present the equation of motion and spectrum of neutral vector boson with MM. Section~\ref{thermo} contains a derivation of the thermodynamical potential and other properties the system, such as Bose-Einstein condensation and self-magnetization. Section~\label{conclusions} is devoted to conclusions.

\section{Equation of motion of a neutral vector boson bearing an MM}
\label{eom}
The Lagrangian density of a charged spin-one particle with MM that moves in a  magnetic field is
\begin{eqnarray}\label{Lagrangian}
  L = \frac{1}{2} U_{\mu\nu}^\dagger (\partial_{\mu} U_{\nu}-\partial_{\nu} U_{\mu}) +
      \frac{1}{2} (\partial_{\mu} U_{\nu}^{\dagger}-\partial_{\nu} U_{\mu}^{\dagger}) U_{\mu\nu} -
      \frac{1}{2} U_{\mu\nu}^{\dagger} U_{\mu\nu} + m^2 U_{\mu}^{\dagger} U_{\mu}\\\nonumber +
      i m \kappa(U_{\mu}^{\dagger} U_{\nu}-U_{\nu}^{\dagger} U_{\mu}) F_{\mu\nu},
\end{eqnarray}

Eq.  (\ref{Lagrangian}) is an extension of the original Proca Lagragian for spin one particles that includes the interaction of the bosons with an external field \cite{PhysRev.131.2326,PhysRevD.89.121701}. The indices $\mu$ and $\nu$ run from 1 to 4, $F^{\mu\nu}$ is the electromagnetic tensor and $U_{\mu\nu}$, $U_{\mu}$ are independent field variables with equations \cite{PhysRev.131.2326}

\begin{equation}\label{fieldeqns}
  \partial_{\mu} U_{\mu\nu}-m^2 U_{\nu}+ 2i \kappa m U_{\mu} F_{\mu\nu}=0,\quad\quad
  U_{\mu\nu} = \partial_{\mu} U_{\nu} - \partial_{\nu} U_{\mu}.
\end{equation}

In the momentum space the field equation is
\begin{equation}
((p_{\mu}^2  + m^2)\delta_{\mu\nu} -p_{\mu} p_{\nu}  - 2  i \kappa m F_{\mu \nu})U_{\mu} = 0.
 \end{equation}

Then, the boson propagator read as
\begin{equation}\label{propagator}
D_{\mu\nu}^{-1}=((p_{\mu}^2  + m^2)\delta_{\mu\nu}-p_{\mu} p_{\nu}  - 2  i \kappa m F_{\mu \nu}).
\end{equation}

Starting from Eqs. (\ref{fieldeqns}) and following the same procedure of Ref. \cite{PhysRev.131.2326} we can obtain the generalized Sakata-Taketani equation for a  six-component wave function
and the  Hamiltonian \cite{PhysRev.131.2326, PhysRevD.89.121701} of the system
\begin{equation}\label{hamiltonian}
H = \sigma_3 m + (\sigma_3 + i \sigma_2) \frac{p^2}{2 m} -
    i \sigma_2 \frac{(\overrightarrow{p}\cdot\overrightarrow{S})^2}{m}
    -(\sigma_3 - i \sigma_2) \kappa \overrightarrow{S} \cdot \overrightarrow{B},
\end{equation}

\noindent where $p=(p_{\perp},p_3)$, $p_{\perp}=p_1^2 + p_{2}^2$, $\sigma_{i}$ are the $2\times2$ Pauli
matrices,
\footnote{
$\begin{array}{ccc}\sigma_1=
 \left(
\begin{array}{cc}0 & 1  \\1 & 0 \end{array}
\right),
& i\sigma_2=
\left(
\begin{array}{cc} 0 & 1  \\ \text{-}1 & 0\end{array}
\right),
&\sigma_3=
\left(
\begin{array}{cc}1&0\\0&\text{-}1\end{array}
\right)
\end{array}$ \\[1pt]}
$S_{i}$ are the $3\times3$ spin-1 matrices in a representation in which $S_3$ is diagonal and $\overrightarrow{S} = \{S_1,S_2,S_3\}$\footnote{
$\begin{array}{ccc} S_1=\frac{1}{\sqrt{2}}
\left( \begin{array}{ccc}
0 & 1& 0\\
1 & 0 & 1\\
0 & 1 & 0
\end{array}\right),
& S_2=\frac{i}{\sqrt{2}}
\left( \begin{array}{ccc}
0 & \text{-}1& 0\\
1 & 0 & \text{-}1\\
0 & 1 & 0
\end{array} \right),
& S_3=
\left( \begin{array}{ccc}
1 & 0& 0\\
0 & 0 & 0\\
0 & 0 & \text{-}1\end{array}\right)\end{array}$}. The magnetic field is considered uniform, constant and in $p_3$ direction $\overrightarrow{B}=B\overrightarrow{e_3}$.

The equations of motion for the momentum $\vec{p}$ and the position $\vec{r}$ can be obtained from:

\begin{eqnarray}
\frac{\partial \vec{p}}{\partial t} = i [H,\vec{p}] \\\nonumber
\frac{\partial \vec{r}}{\partial t} = i [H,\vec{r}] \nonumber
\end{eqnarray}

\noindent and they read:

\begin{equation} \label{motion}
 \frac{\partial \vec{p}}{\partial t}=\vec{0},
 \end{equation}

\begin{equation} \label{motionr}
m  \frac{\partial \vec{r}}{\partial t}= (\sigma_3 - i \sigma_2) \vec{p} + i \sigma_2 [\vec{S}, \vec{p}. \vec{S}].
 \end{equation}

\noindent Here $[a,b] = ab-ba$ is the commutator of $a$ and $b$.

From Eq.(\ref{motion}) follows that the neutral bosons move freely in the direction parallel to the field as well as in the perpendicular one. This is a difference with respect to the charged vector boson case, in which the perpendicular component of the momentum is quantized \cite{Elizabeth}.

The eigenvalues of Eq.(\ref{hamiltonian}) are
\begin{equation}
E(p_{\perp},p_3, B)=\sqrt{p_3^2+p_{\perp}^2+m^2-2\kappa s B\sqrt{p_{\perp}^2+m^2}},\label{spectrum}
\end{equation}
\noindent where $s$ are the spin eigenvalues $s=0, \pm 1$. The effective rest mass of the neutral spin one boson ($s=1$ and $p_3=p_{\perp}=0$) has the form
\begin{equation}
E(0, B)=\sqrt{m^2-2\kappa B m}.\label{massrest}
\end{equation}
\noindent  Eq. (\ref{massrest}) shows a decreasing behavior when the magnetic field increase and a critical magnetic field is obtained at $B_c=m/2\kappa$. As $m=2m_e$ and $\kappa=2\mu_B$, $B_c=m_e^2/e=4.41 \times 10^{13}~G$ which is the Schwinger critical field. We can define the effective magnetic moment as
\begin{equation}
d=-\frac{\partial E}{\partial B}=\frac{\kappa m}{\sqrt{m^2-2m\kappa B}},\label{magmoment}
\end{equation}
\noindent as we can see from (\ref{magmoment}) the system has a paramagnetic behavior because $d>0$ and increase with the increasing of the magnetic field. It has a singularity when $B=B_c$.

\section{Thermodynamical properties of a boson gas}
\label{thermo}
The general expression for the thermodynamical potential of a neutral vector boson gas has the form
\begin{equation}\label{Grand-Potential-4}
\Omega(B,\mu,T)= \frac{1}{\beta}\left[\sum_{p_4}\int\limits_{-\infty}^{\infty}\frac{p_{\perp}dp_{\perp}dp_3}{(2\pi)^2} \ln \det D^{-1}(\overline{p}^*)\right].
\end{equation}
Here $D^{-1}(\overline{p}^*)$ is the neutral boson  propagator given by Eq. (\ref{propagator}),  $\beta = 1/T$  denotes the inverse temperature, $\mu$ the boson chemical potential and ${\overline{p}}^*=(ip^{4}-\mu,0,p_{\perp},p_{3})$.
After doing the Matsubara sum the expression (\ref{Grand-Potential-4}) becomes
\begin{equation}
\Omega(B,\mu,T)= \frac{1}{\beta}\left[\int\limits_{0}^{\infty}\frac{p_{\perp}dp_{\perp}dp_3}{(2\pi)^2}[ \ln f^{-}f^{+} +E/\beta] \right].
\end{equation}
\noindent The logarithm terms are called the statistical contribution of bosons/antibosons,  they depend on $B, T$ and $\mu$. The term $E/\beta$ is only B-dependent and is called the vacuum term. For magnetic field below  $B_c$ the leading term of thermodynamical potential is the statistical one so we neglect the vacuum term in what follows.  The functions $f^{\pm}$ are
\begin{equation}\label{distribution}
f^{\pm}=1-e^{-(E\mp \mu)\beta}.
\end{equation}
The density of bosons can be calculated as
\begin{equation}\label{density}
N=-\frac{\partial \Omega}{\partial B} =\int\limits_{-\infty}^{\infty}\frac{p_{\perp}dp_{\perp}dp_3}{(2\pi)^2}(n^{+}-n^{-}),
\end{equation}
\noindent where $n^{\pm}$ are the Bose-Einstein distribution of bosons and antibosons as
\begin{equation}\label{distribution}
n^{\pm}=\frac{1}{e^{(E\mp \mu)\beta}-1}.
\end{equation}
We focus in Astrophysical scenarios, so in our approach we consider the "cold" degenerate boson gas where $\mu>T$ which correspond to the degenerate limit and antibosons are neglected. Besides,  we will concentrate in the study of the ground state phase determined by Eq(\ref{massrest}). We have also consider that our system is one dimensional ($p_{\perp}=0$), therefore the boson distribution is only a function of $p_3$ and the integral for particle density has the form
\begin{equation}\label{densityD}
N=\int\limits_{-\infty}^{\infty}\frac{dp_3}{2\pi} n^{+}(p_3).
\end{equation}

For  $T\rightarrow 0$,  $\mu \rightarrow E(0,B)$ the particle distribution Eq.(\ref{distribution}) concentrate around $p_3=0$. In this limit the particle density Eq.(\ref{densityD})  can be approximated as
\begin{eqnarray}
N \simeq \frac{1}{\pi \beta} \int_{0}^{p_0} \frac{dp_3}{\sqrt{p_3^2+m^2-2 \kappa B m}-\mu}
  \simeq \frac{1}{\pi \beta} \int_{0}^{p_0} \frac{2 E(0,B) dp_3}{p_3^2+ E(0,B)^2-\mu^2},\nonumber
\end{eqnarray}
\begin{eqnarray}\label{N}
N \simeq \frac{1}{2 \beta} \sqrt{\frac{2 E(0,B)}{-\mu^\prime}}.
\end{eqnarray}
 Here we have defined $\mu^\prime = \mu - E(0,B)$ and $p_0$ is some characteristic moment such as $p_0 >> \sqrt{-2 E(0,B) \mu^\prime}$. We expect that $\mu^{\prime} \rightarrow 0$ when $T \rightarrow 0$, because this is equivalent to
the condensation condition $\mu \rightarrow E(0,B)$. An expression for $\mu^\prime(T)$ can be obtained from Eq. (\ref{N})
\begin{equation}\label{miuprima}
\mu^\prime = -\frac{E(0,B) T^2}{2 N^2}
\end{equation}
From Eq.(\ref{miuprima}) one can see that $\mu^\prime$ is a decreasing function of T, as occurs in the usual Bose-Einstein condensation.
By the use of Eq.(\ref{miuprima}) we can approximate $n^{+}(p_3)$ in a vicinity of $p_3 =0$ as
\begin{eqnarray}
n^+(p_3) \simeq \frac{2 E(0,B) T}{p^2 +E(0,B)^2 - \mu^2} \simeq \frac{2 E(0,B) T}{p^2 -2 E(0,B) \mu^\prime} \\\nonumber
\end{eqnarray}
\begin{eqnarray}\label{nmas}
n^+(p_3) \simeq 2 N \frac{\gamma}{p_{3}^2 + \gamma^2}
\end{eqnarray}
\noindent where $\gamma = \sqrt{- 2 E(0,B) \mu^\prime}$. Eq. (\ref{nmas}) means that for $p_3 \approx 0$, $n^+(p_3)$ can be approximated by a
Cauchy distribution centered in $p_3=0$. Now the equivalent to the $T \rightarrow 0$ limit is $\gamma \rightarrow 0$
\begin{equation}
\lim_{\gamma\rightarrow 0} n^+(p_3) = 2 N \delta(p_3),
\end{equation}
\noindent and finally we have the following expression for the particle density $N$ in the vicinity of $p_3=0$
\begin{equation}\label{densityTcero}
N\simeq \frac{1}{2} \lim_{\gamma\rightarrow 0}\int_{-\infty}^{\infty} 2 N \frac{\gamma}{p_{3}^2 + \gamma^2} dp_3 = N \int_{-\infty}^{\infty} \delta(p_3)dp_3.
\end{equation}
In the limit $T \rightarrow 0$  Eq. (\ref{densityTcero}) shows that all the bosons  fall in the condensate since  the total density of bosons depends on the $\delta(p_3)$, but there are not a critical temperature for condensation to start and it is reached essentially  in the ground state. This condensation is called diffuse \cite{ROJAS1996148,PEREZROJAS2000,Khalilov2001}.
	
\begin{figure}[t]
\centerline{
\includegraphics[width=0.5\linewidth]{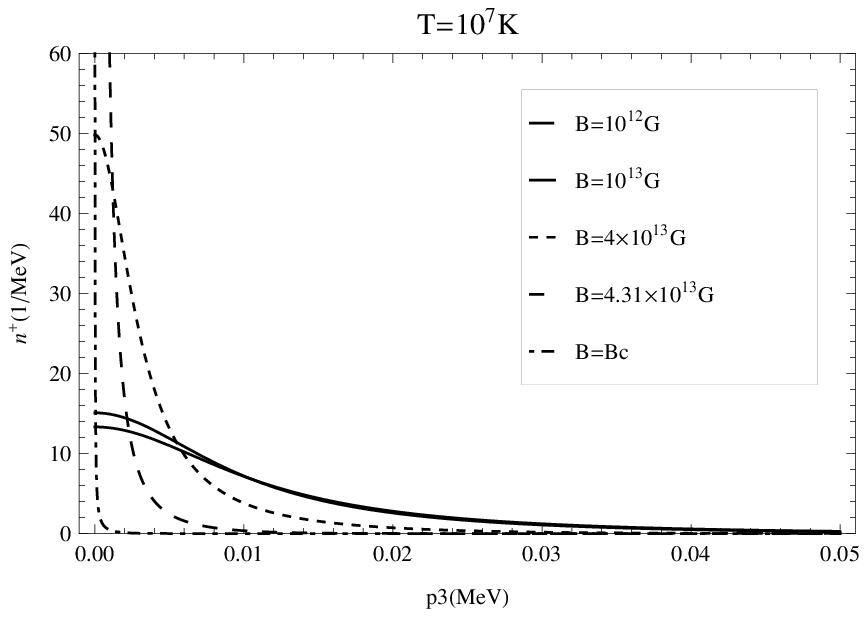}
\includegraphics[width=0.5\linewidth]{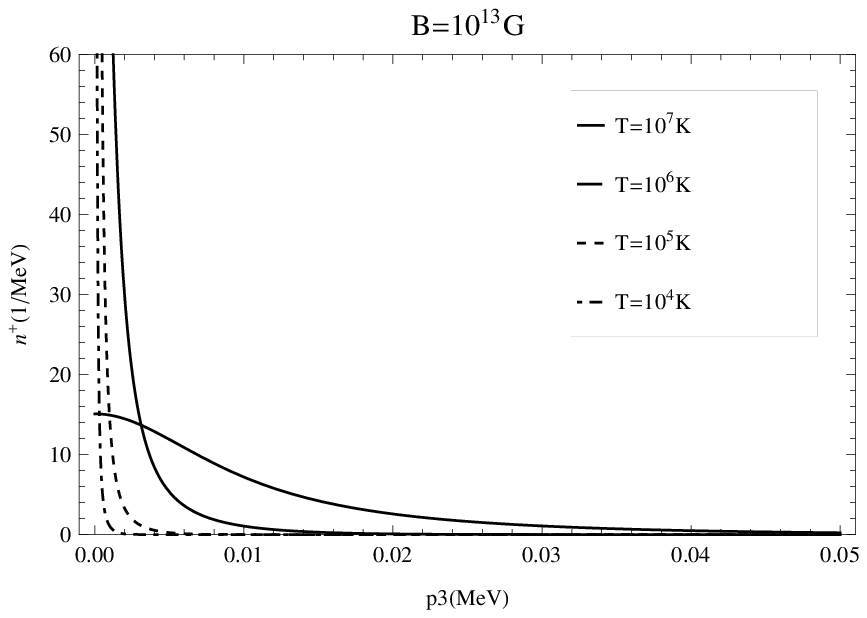}
		\vspace*{8pt}}	
\caption{\label{fig1}  Density of bosons as a function of temperature for several values of magnetic field(left plot) and density of boson as a function of $p_3$ for different values of temperatures  (right plot). In both graphics $N=10^{34} cm^{-3}$.}
\end{figure}

Fig 1 shows the $\delta$-behavior around $p_3\sim 0$  of density of bosons Eq.(\ref{density}).
 In left panel of Fig 1 is illustrated the density of bosons as a function of $p_3$ for a fixed value of temperature $T=10^7 K$ and several values of the magnetic field. The curves move to the left when the magnetic field grows.
Right panel of Fig 1 shows the density of bosons as a function of $p_3$ for a fixed value of the magnetic field $B=10^{13}~G$ and  different values of the temperatures.  The curves shift to the left with the decreasing of the temperature.

Eq. (\ref{N}) also allows us to compute the thermodynamical potential and the magnetization as
\begin{equation}\label{omega}
\Omega = \frac{1}{ \beta} \sqrt{E(0,B)^2-\mu^2},
\end{equation}
\begin{equation}\label{magnetization}
M=-\frac{\partial \Omega}{\partial B}= \frac{\kappa m}{E(0,B)} N = N d.
\end{equation}
In Fig.~2 left the magnetization (Eq.(\ref{magnetization}))  is plotted as a function of magnetic field for several values of the particle density.

\begin{figure}[h]
\centerline{
\includegraphics[width=0.52\linewidth]{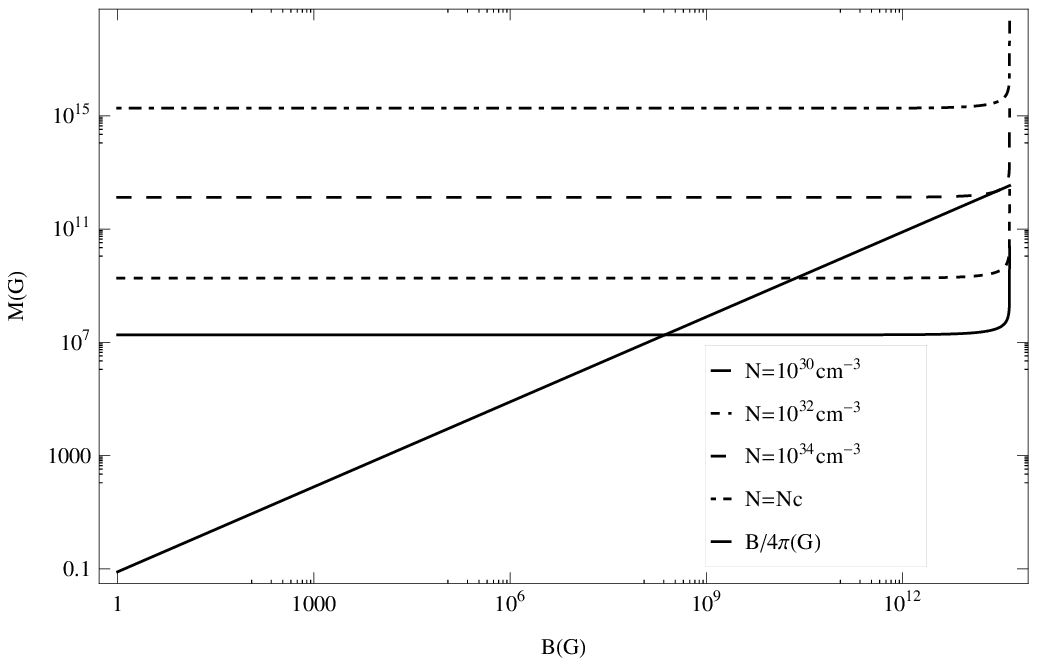}
\includegraphics[width=0.5\linewidth]{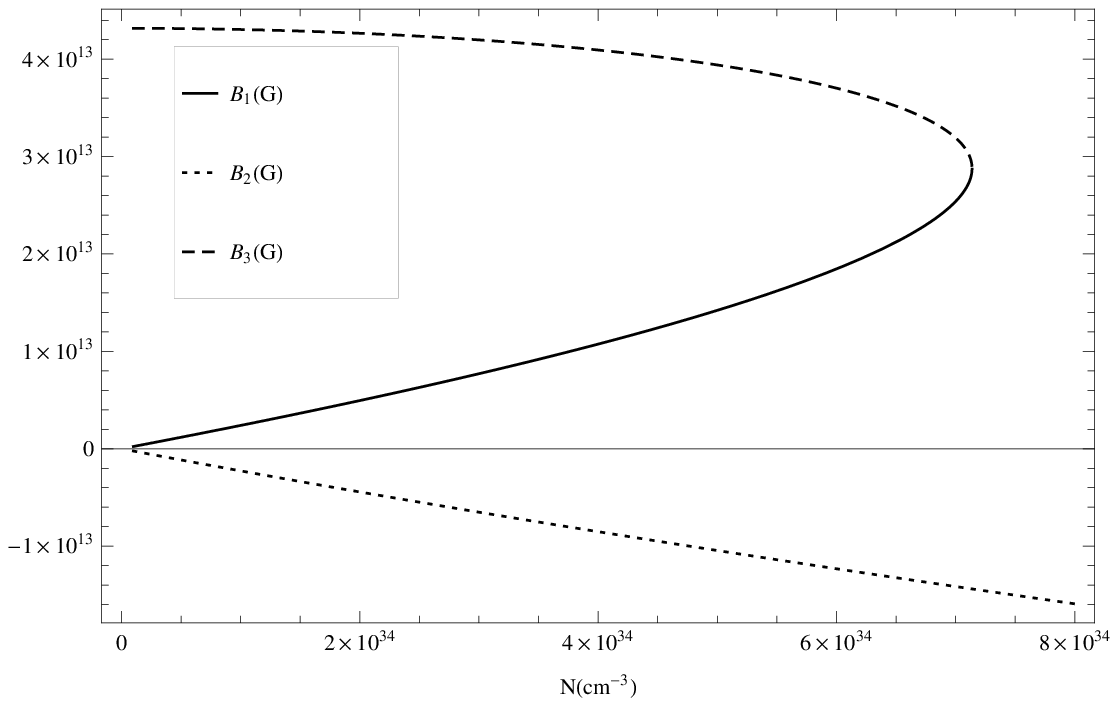}
		\vspace*{8pt}}	
\caption{\label{fig2} Magnetization as a function of magnetic field. We have also plotted the $B/4\pi$ for T=0 (left plot) and we have depicted the solution of self-magnetization equation as a function of particle density $N$ (right plot).}
\end{figure}

 We are interested in search if the system reaches the self-magnetization condition considering $H = B-4 \pi M$ with $H=0$ and solving the self-consistent equation $B=4\pi M$. This is a cubic equation due to the non linear dependency of the magnetization on the field.  In Fig ~2 right its three solutions have been plotted but only one of them is physically meaningful. For one of the roots, the magnetic field is negative (see dotted line), while for another, it decreases with the increasing density, reaching $B_c$ when $N$ goes to zero (dashed line). These solutions implies that the magnetization also decreases with $N$ and it is contrary to Eq. (\ref{magnetization}). Both roots are nonphysical, and hence, they must be discarded. Therefore, the only admissible solution of the self-magnetization equation is given by the solid line. The solution becomes complex for densities higher than $N_c = 7.14 \times 10^{34}cm^{-3}$ which corresponds to a value of the magnetic field of $2/3\times B_c$. Being the magnetization a positive quantity, the points of the solid line are the values of the self-maintained magnetic field and shows the ferromagnetic behavior of the boson gas.



\section{Conclusions}\label{conclusions}

We studied the equation of motion of a neutral vector boson  with MM in presence of magnetic field. The spectrum is calculated starting from the generalized Sakata-Taketani hamiltonian\cite{PhysRev.131.2326,PhysRevD.89.121701}.

The thermodynamical potential of the boson gas in the ground state is studied. The gas exhibits a ``difuse'' Bose Einstein condensation characterized  by the non existence of critical temperature for the phase transition and the growing population of the ground state when the magnetic field increases or  the temperature  decreases.

The magnetization of the system is also calculated for the ground state, it is a positive quantity and grows with field up to $B_c$ where it diverges.  For particle densities under a critical value the system can  maintain the magnetic field because the self-magnetization equation has a positive solution.  This phenomenon appears for values of densities and magnetic fields that are typical of compact objects and it could have relevance in modelling jets as well as a mechanism to sustain the strong magnetic field in compact objects. These models deserve a separate work which is in progress.


\end{document}